\documentclass[nofootinbib,reprint,amssymb,prd,aps,superscriptaddress]{revtex4-1}
\usepackage{graphicx}
\usepackage{xcolor,amsmath}
\usepackage{multirow}
\usepackage{comment}
\usepackage{cases}
\usepackage{bigints}
\usepackage{enumitem}
\usepackage{hyperref}
\hypersetup{colorlinks, linkcolor={teal},citecolor={blue},urlcolor={blue}}

\newcommand{\dd}{{\rm d}}

\begin{document}

\title{Observational constraints on nonlocal black holes via gravitational lensing}

\author{Rocco D'Agostino}
\email{rocco.dagostino@inaf.it}
\affiliation{INAF -- Osservatorio Astronomico di Roma, Via Frascati 33, 00078 Monte Porzio Catone, Italy}
\affiliation{INFN, Sezione di Roma 1, Piazzale Aldo Moro 2, 00185 Roma, Italy}

\author{Vittorio~De~Falco}
\email{v.defalco@ssmeridionale.it}
\affiliation{Ministero dell'Istruzione e del Merito (M.I.M., ex M.I.U.R.), Italy}

\begin{abstract}
In this paper, we study the gravitational lensing around the static and spherically symmetric DD black holes, which we recently derived as perturbations of the Schwarzschild geometry within the revised Deser-Woodard theory of nonlocal gravity. We first present general analytical expressions for the deflection angle in both weak- and strong-deflection limits, explicitly relating them to the nonlocal corrections to Schwarzschild spacetime. Subsequently, we analyze lensing observables, such as the post-Newtonian effects and the black hole shadow, to constrain the DD black hole parameter space using current observational bounds. Finally, we perform a joint statistical analysis based on the Fisher information matrix, combining these findings with our previously obtained constraints from quasinormal modes. Our results indicate consistency with general relativity at the $1.13\sigma$ level. This work provides a first assessment of the DD parameter space and offers new insights to probe deviations from Einstein's gravity in view of future larger datasets.
\end{abstract}

\maketitle

\section{Introduction}
Black holes (BHs) represent the most fascinating predictions of relativistic gravity and provide a unique physical laboratory for testing the behavior of spacetime in the strong-field regime \cite{Will:2014kxa,Berti:2015itd}. The increasing precision of astrophysical observations, ranging from gravitational-wave (GW) detections to horizon-scale imaging, has opened unprecedented opportunities to probe General Relativity (GR) in the vicinity of compact objects \cite{LIGOScientific:2016lio,EventHorizonTelescope:2019dse,EventHorizonTelescope:2022xqj}. 

Despite its experimental success, GR faces both conceptual and observational challenges. On the theoretical side, the occurrence of spacetime singularities and the absence of a consistent quantum theory of gravity point toward the need for going beyond Einstein's framework \cite{Ashtekar:2004eh,Kiefer:2013jqa}. From an observational perspective, the standard cosmological paradigm based on GR requires the introduction of a cosmological constant and cold dark matter to account for the present accelerated expansion of the Universe \cite{SupernovaSearchTeam:1998fmf,SupernovaCosmologyProject:1998vns,Planck:2018vyg}. However, these ingredients introduce fine-tuning and coincidence problems \cite{Weinberg:1988cp,Zlatev:1998tr,Padmanabhan:2002ji} and leave the fundamental nature of dark energy and dark matter unexplained \cite{Peebles:2002gy,Bertone:2004pz,Feng:2010gw,DAgostino:2019wko,DAgostino:2022fcx}. Such considerations provide valid motivations for exploring modified theories of gravity \cite{Carroll:2003wy,Sotiriou:2008rp,DeFelice:2010aj,Clifton:2011jh,Joyce:2014kja,Nojiri:2017ncd,Capozziello:2019cav,DAgostino:2019hvh,DAgostino:2020dhv,CANTATA:2021asi,DAgostino:2022tdk,DAgostino:2024sgm,DAgostino:2025kme}, whose deviations from GR may become relevant either on cosmological scales or in strong-field environments, while remaining compatible with local tests.

Among the proposed extensions of GR, nonlocal gravity offers a mathematically well-motivated framework in which infrared modifications arise through operators involving inverse powers of the d’Alembertian \cite{Biswas:2005qr,Deser:2007jk,Barvinsky:2011rk,Maggiore:2013mea,Kehagias:2014sda,Calcagni:2014vxa,Deser:2019lmm}. These constructions preserve general covariance and can be formulated, at the perturbative level, without introducing additional propagating ghosts.
In particular, the revised Deser-Woodard model~\cite{Deser:2019lmm} has emerged as a viable alternative to the standard cosmological scenario, due to its ability to reproduce the observed cosmic evolution without fine-tuning, while naturally implementing a screening mechanism that ensures consistency with Solar System tests \cite{Capozziello:2023ccw}. This description has also been used to investigate large-scale structure formation and nonsingular bouncing cosmologies \cite{Ding:2019rlp,Chen:2019wlu,Jackson:2023faq}.

Building on this framework, we have recently demonstrated the existence of new traversable wormhole solutions~\cite{DAgostino:2025sta}, and a family of static and spherically symmetric spacetimes with asymptotically flat behavior, known as DD BHs~\cite{DAgostino:2025wgl}. These solutions are characterized by a small deformation parameter around the Schwarzschild spacetime and by a real exponent controlling the strength and radial profile of the nonlocal corrections. The DD BHs have later been employed to investigate quasinormal modes (QNMs), serving both as a theoretical framework for deriving axial and polar gravitational perturbations~\cite{DAgostino:2026wln} and as a basis for observational analyses aimed at constraining the associated parameter space~\cite{DAgostino:2025yej}. 

In this observational context, a methodology has recently been introduced in the GW domain to distinguish between the Schwarzschild and the DD geometries by exploiting the dynamics of periodic orbits and the related gravitational waveforms emitted by spinning particles~\cite{Bravo-Gaete:2026com}. Further investigations have explored the weak-field regime through Solar System experiments~\cite{Liu:2025cpp}.

A first assessment of the gravitational lensing signatures of the DD BHs has been conducted in Ref.~\cite{Li:2026zbo}.
Gravitational lensing, in particular, provides a sensitive probe of nonlocal features. Indeed, light propagation is entirely determined by the spacetime geometry and is largely insensitive to the detailed astrophysical properties of the lens, the source, or the surrounding environment~\cite{Virbhadra:1999nm,Perlick:2004tq}. Several studies have developed polynomial and semi-polynomial approximations to model photon trajectories and related observables~\cite{Beloborodov:2002mr,Poutanen:2006hw,DeFalco:2016yox,LaPlaca:2019rjz,Poutanen:2019tcd,Bakala:2023fnc}. 
Although these approaches are timely and insightful, a systematic analysis of light propagation in the weak- and strong-field regimes is more suitable for tightly constraining the allowed BH parameter space.

In the weak-deflection limit (WDL), relevant for photons with large impact parameters far from the BH, the bending angle encodes information about the asymptotic structure of the metric and hence about the far-field deviations from GR~\cite{Kaiser:1991qi,Bartelmann:1999yn,Keeton:2005jd,Amendola:2007rr}. 
Recent progress in the WDL has been achieved through alternative geometrical approaches and generalizations that account for dispersive media and massive particle propagation, providing a unified treatment of different lensing scenarios~\cite{Crisnejo:2018uyn,Crisnejo:2019ril}.

In the strong-deflection limit (SDL), associated with trajectories near the photon sphere, photons can undergo multiple loops around the BH before escaping, giving rise to relativistic images~\cite{Treu:2010uj,Cunha:2018acu,Bozza:2002zj}. This regime is highly sensitive to the near-horizon geometry, making it a direct probe of modifications that affect unstable circular null orbits~\cite{Cardoso:2008bp,Virbhadra:2008ws}.
The analytical formalism for SDL has been extended 
beyond the original static and spherically symmetric treatment to cover a wider range of astrophysically relevant configurations, including arbitrary source distances~\cite{Bozza:2006nm,Bozza:2007gt}.
More recent works have generalized the formalism to interferometric signatures of relativistic images~\cite{Aratore:2021usi,Feleppa:2025ejh},
massive particles and inhomogeneous plasma environments~\cite{Feleppa:2024vdk,Feleppa:2024kio}.

In this paper, we revisit and refine the study presented in Ref.~\cite{Li:2026zbo}, clarifying several aspects of the analytical derivation of light deflection and employing current astrophysical measurements to infer observational constraints for the DD BH parameters. We show that this framework provides a largely theory-independent observational test, in which distinct light-propagation regimes probe complementary regions of the spacetime~\cite{Frittelli:1999yf,Ezquiaga:2020dao}. Owing to their smooth reduction to the GR limit, DD BHs provide a controlled setting for investigating lensing observables in both the WDL and SDL regimes, thereby enabling a systematic exploration of how nonlocal modifications of gravity manifest in observable lensing effects. 

The work is organized as follows. In Sec.~\ref{sec:DW-nonlocal}, we review the essentials of the DD BH solution in nonlocal gravity. In Sec.~\ref{sec:GL}, we analyze photon motion and gravitational lensing, deriving the deflection angle in both the WDL and SDL. Applications to lensing observables are discussed in Sec.~\ref{sec:applications}, where we combine our findings with existing constraints in the literature to statistically quantify deviations from GR. Finally, in Sec.~\ref{sec:conclusions}, we summarize our results and outline possible future developments.

In this paper, we adopt units with $G=c=1$ and set the BH mass to unity, unless otherwise specified.

\section{The DD black holes}
\label{sec:DW-nonlocal}

The DD BHs are static and spherically symmetric solutions~\cite{DAgostino:2025wgl} of the revised Deser-Woodard model of nonlocal gravity, whose action is given by \cite{Deser:2019lmm} 
\begin{equation} \label{eq:nonlocal-action}
S=\dfrac{1}{16\pi}\int \sqrt{-g}\, R\left[1+f(Y)\right]\dd^4 x \,,
\end{equation}
where $f(Y)$ is the distortion function containing the nonlocal effects, $R$ is the Ricci scalar, and $g$ stays for the determinant of the metric tensor, $g_{\mu\nu}$.

The dynamics of this model can be studied through a localized form involving four independent auxiliary scalar fields governed by the following equations of motion:
\begin{subequations}
\begin{align}
    \Box X&=R\,, \label{eq:X} \\ 
    \Box Y&=g^{\mu\nu}\partial_\mu X\partial_\nu X\,, \label{eq:Y}\\
    \Box U&=-2\nabla_\mu (V\nabla^\mu X)\,, \label{eq:U} \\
    \Box V&=R f_{,Y}\,,\label{eq:V}
\end{align}
\end{subequations}
where $\Box\equiv \nabla_\mu \nabla^\mu$ is the d'Alembert operator.

The vacuum field equations are obtained by varying the action \eqref{eq:nonlocal-action} with respect to $g_{\mu\nu}$, leading to
\begin{align}
\left(G_{\mu\nu}+g_{\mu\nu}\Box-\nabla_\mu \nabla_\nu\right)W+\mathcal{K}_{(\mu\nu)}-\frac{1}{2}g_{\mu\nu}g^{\alpha\beta}\mathcal{K}_{\alpha\beta} =0\,,
\label{eq:FE}
\end{align}
where $G_{\mu\nu}\equiv R_{\mu\nu}-\frac{1}{2}g_{\mu\nu}R$ is the Einstein tensor, and $W\equiv 1+U+f(Y)$. Moreover,
\begin{equation}
\mathcal{K}_{\mu\nu}\equiv \partial_\mu X\partial_\nu U +\partial_\mu Y \partial_\nu V+V\partial_\mu X \partial_\nu X \,,
\label{eq:Kmunu}
\end{equation}
and $\mathcal{K}_{(\mu\nu)}\equiv (\mathcal{K}_{\mu\nu}+\mathcal{K}_{\nu\mu})/2$.

We assume a static and spherically symmetric metric of the general form in spherical-like coordinates
\begin{equation}
\dd s^2=-A(r) \dd t^2+\frac{\dd r^2}{B(r)} +C(r)(\dd \theta^2+\sin^2\theta\, \dd\varphi^2)\,,
    \label{eq:general-metric}
\end{equation}
where $A(r),B(r),$ and $C(r)$ are positive functions outside the event horizon satisfying $\{A(r),B(r)\}\to 1$ as $r\to\infty$. 

As demonstrated in Ref.~\cite{DAgostino:2025wgl}, it is possible to search for BH solutions of Eq.~\eqref{eq:FE} by recasting the field equations in the tetrad frame linked to a static observer in spacetime~\eqref{eq:general-metric}. In this way, we can determine the following family of solutions:
\begin{subequations}\label{eq:BH_sol}
\begin{align}
    A(r)&=1-\frac{2}{r}-\frac{\xi }{r^k}\,, \label{eq:BH_solA}\\
    B(r)^{-1}&=1-\frac{2}{r}+\frac{\xi}{3^{k} r^{k+1}(r-3)^2} \left\{3^k r \Big{[}k (r-3) (r-2)\right.\notag\\
    &\left.\quad+4 r-9\Big{]}-3 (r-2) (2 r-3) r^k\right\},\label{eq:BH_solB}
    \\
    C(r)&=r^2\,,\label{eq:BH_solC}
\end{align}    
\end{subequations}
where $k>1$ is a real constant and $0<\xi \ll 1$ is a small parameter encoding first-order perturbations around the Schwarzschild metric.
The analytical expressions for the auxiliary scalar fields $\{X(r),Y(r),U(r),V(r)\}$, as well as for the distortion function $f(Y)$, can be found in Ref.~\cite{DAgostino:2025wgl}.

The obtained solution is asymptotically flat, contains the essential singularity in $r=0$, and is regular outside the event horizon, which is given by
\begin{equation}
r_H=2+\frac{\xi}{2^{k-1}}\,.
\label{eq:rH}
\end{equation}
An important geometrical quantity, useful for the subsequent calculations, is the photon sphere
\begin{align}
r_m&=3 \left[1+\xi \left(\frac{k+2}{2\times 3^k} \right)\right],
\end{align}
which is the solution of the following differential equation
\begin{equation} 
A_m C_m'-A'_mC_m =0\,.    
\label{eq-ps}
\end{equation}
Here, the prime denotes differentiation with respect to the radial coordinate $r$, and all quantities with the subscript $m$ are evaluated at $r=r_m$.

\section{Gravitational lensing}
\label{sec:GL}

\begin{figure}
    \centering
    \includegraphics[width=3.5in]{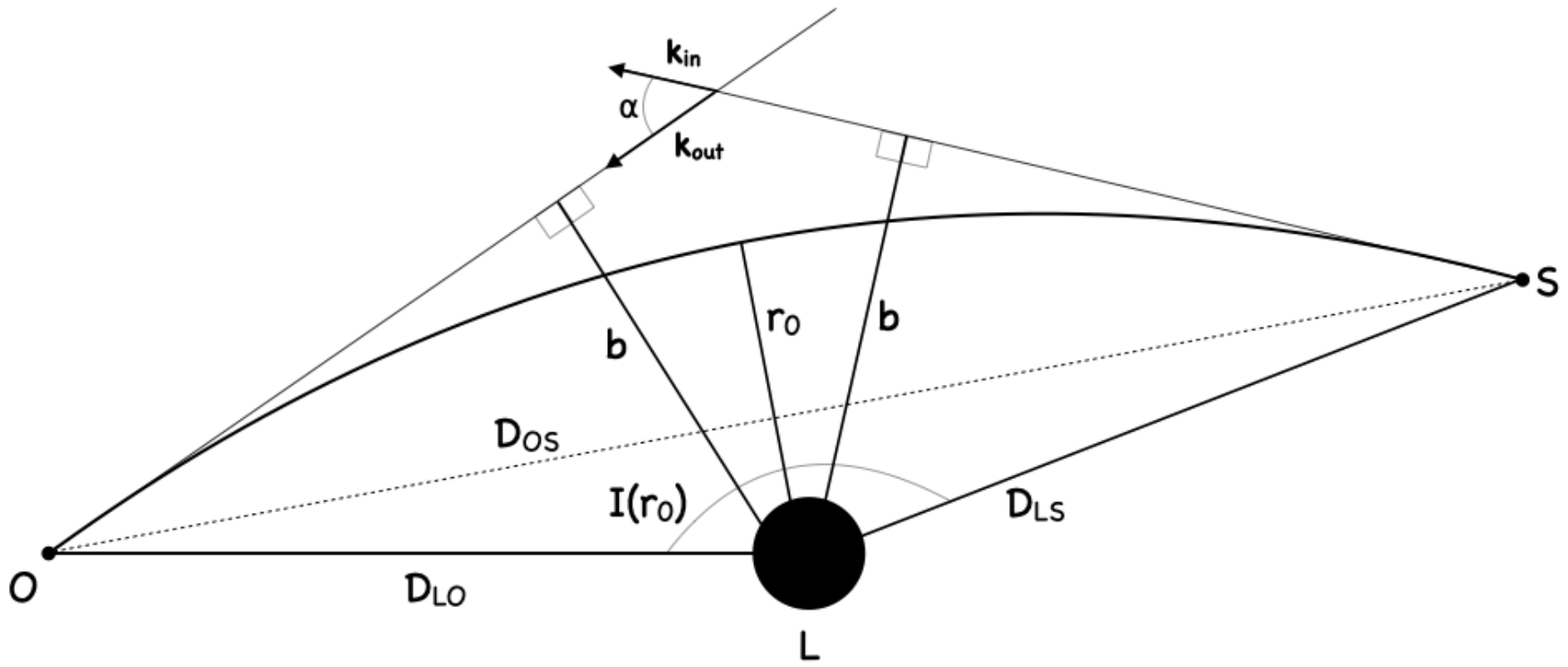}
    \caption{Geometry of the gravitational lensing configuration framed in a static and spherically symmetric spacetime \eqref{eq:general-metric}. }
    \label{fig:Lens_configuration}
\end{figure}

In this section, we analyze the gravitational lensing of photon trajectories passing close to a BH, described by the generic metric~\eqref{eq:general-metric}. 
The photon is emitted by a source $S$, located at a distance $D_{\rm LS}$ from the BH lens $L$, and reaches a static observer $O$ situated at a distance $D_{\rm LO}$ from the BH. Therefore, the separation between $S$ and $O$ is $D_{\rm OS}$ (see Fig.~\ref{fig:Lens_configuration} for a schematic representation).

Photon motion is described by null geodesics ($\dd s^2=0$) that, without loss of generality, can be set in the equatorial plane $\theta=\pi/2$ by virtue of spherical symmetry. In the spacetime~\eqref{eq:general-metric}, one can always identify two conserved quantities along null geodesics:
\begin{align}\label{eq:CQ}
E = A(r)\,\dot{t}\,,\quad 
L = C(r) \dot{\varphi}\,,
\end{align}
where the dots denote derivatives with respect to an affine parameter $\lambda$ parameterizing the photon trajectory. In asymptotically flat spacetime, $E$ and $L$ represent the photon energy and angular momentum, respectively, measured by static observers at infinity. 
The photon impact parameter is defined as $b \equiv L/E$, which corresponds to the perpendicular distance between the center of the lens and the asymptotic, undeflected (straight) photon trajectories with unit direction $\boldsymbol{\hat{k}}_{\rm in}$ and $\boldsymbol{\hat{k}}_{\rm out}$ (see Fig.~\ref{fig:Lens_configuration}).

We now relate the impact parameter $b$ to the distance of closest approach (or periastron radius) $r_0$ of the photon trajectory to the BH. Imposing the null condition yields
\begin{equation}
0 = -A(r)\dot{t}^2 + B(r)\dot{r}^2 + C(r)\dot{\varphi}^2 .
\end{equation}
Substituting the conserved quantities~\eqref{eq:CQ}, one obtains
\begin{equation}\label{eq:raddot}
\dot{r}^2 = \frac{E^2}{A(r)B(r)}
\left[
1 - \frac{A(r)b^2}{C(r)}
\right].
\end{equation}
The distance of closest approach $r_0$ is determined by the turning point condition $\dot{r}=0$, which yields
\begin{equation}\label{eq:impact}
    b\equiv b(r_0)=\sqrt{\frac{C_0}{A_0}}\,,
\end{equation}
where the subscript $0$ means evaluation at $r=r_0$.
We recall that $r_0\in[r_m,\infty)$, since for $r_0<r_m$ photons are captured by the BH, and no turning point exists. Equivalently, 
$b_m$ is the critical impact parameter separating stable orbits ($b>b_m$) from unstable orbits ($b<b_m$), while for $b=b_m$ the photon trajectory asymptotically approaches the photon sphere at $r_m$. Its expression is~\cite{DAgostino:2025wgl}
\begin{equation}
b_m=3\sqrt{3}+\dfrac{3^{\frac{5}{2}-k}}{2} \xi\,.  
\end{equation}

A photon emitted from $S$ at angle $\varphi_{\rm S}$, reaching $r_0$, and escaping far from the source to $O$ at angle $\varphi_O$ experiences a total azimuthal variation\footnote{To easily define and compute Eq.~\eqref{eq:DA}, we make two assumptions: (1) $S$ and $O$ are both located at infinity; (2) the photon trajectory is symmetric with respect to $r_0$.}
\begin{align}\label{eq:TAA}
I(r_0)&\equiv\int_{\varphi_S}^{\varphi_O}\dd\varphi=2\int_{r_0}^\infty\dd r\,\dfrac{\dot{\varphi}}{\dot{r}}\notag\\
&=2 \int_{r_0}^{\infty} \dd r \left\{\frac{C(r)}{B(r)}\left[\frac{C(r)}{A(r) b^2}-1\right]\right\}^{-1/2},
\end{align}
where we have used Eqs.~\eqref{eq:CQ} and \eqref{eq:raddot}.

In the absence of a gravitational source, the photon trajectory coincides with the straight line $SO$ shown in Fig.~\ref{fig:Lens_configuration}, which subtends an angle $\pi$. Instead, the presence of a BH creates a deflection (or light-bending) angle $\alpha(r_0)$, defined as the angle formed between the unit vectors $\boldsymbol{\hat{k}}_{\rm in}$ and $\boldsymbol{\hat{k}}_{\rm out}$. Therefore, $\alpha(r_0)$ is defined as the deviation of the total azimuthal variation \eqref{eq:TAA} from a straight line:
\begin{equation}\label{eq:DA}
\alpha\left(r_0\right)=I\left(r_0\right)-\pi\,.
\end{equation}

For most theories of gravity, the integration of the deflection angle can be easily computed numerically. However, in the following, we derive the analytical behaviors of Eq. \eqref{eq:DA} in both the WDL and the SDL. These regimes are theoretically well established for spacetimes of the form~\eqref{eq:general-metric}, and the coefficients appearing in the corresponding approximations are naturally associated with physical observables, which can then be directly compared with astrophysical data.

\subsection{Weak-field regime}
\label{sec:WDL}

We start with the WDL, corresponding to $\alpha(r_0)\ll1$, obtained by taking the limit $1/r_0\rightarrow 0$ in Eq.~\eqref{eq:DA}. 
Under the change of variable $z=r_0/r$, Eq.~\eqref{eq:TAA} becomes
\begin{equation}\label{eq:INT-WD}
I\left(r_0\right) = 2r_0\int_{0}^{1} \frac{\dd z}{z^2} \left\{\frac{C(r_0/z)}{B(r_0/z)}\left[\frac{C(r_0/z)}{A(r_0/z) b^2(r_0)}-1\right]\right\}^{-1/2}.
\end{equation}
Using Eq.~\eqref{eq:impact}, the integrand of Eq.~\eqref{eq:INT-WD} can be asymptotically expanded as a power series in $1/r_0$. 
This allows the integral to be computed straightforwardly order by order, yielding
\begin{align} \label{eq:alpha-WDL}
&\alpha(r_0)\!=\!\left(\!4+\dfrac{2\xi}{3^{k-1}}\!\right)\!\dfrac{1}{r_0}-\!\left[4-\dfrac{15 \pi }{4}+\!\left(\!\dfrac{8-15 \pi}{4}\!\right)\!\dfrac{\xi}{3^{k-1}}\right]\!\dfrac{1}{r_0^2}\notag\\
&+\!\left[\frac{122}{3}-\frac{15 \pi }{2}+\left(\dfrac{\!42-5 \pi}{2} \!\right)\!\dfrac{3\xi}{3^{k-1}} \right]\!\dfrac{1}{r_0^3}+\mathcal{O}\!\left(\dfrac{1}{r_0^4}\right)\!.
\end{align}

The above series can also be expressed in terms of the impact parameter $b$. Indeed, using Eq.~\eqref{eq:impact}, one can first expand it in powers of $1/r_0$ and then invert the resulting series to obtain the following expansion
\begin{equation}\label{eq:bexp}
        \frac{1}{r_0}=\sum_{n=1}^\infty a_n(k,\xi) \dfrac{1}{b^n}\,,
\end{equation}
where the coefficients $a_n(k,\xi)$ cannot be determined in closed form, but can be computed once a specific value of $k$ is fixed, while $\xi$ can be kept general. 

Substituting Eq.~\eqref{eq:bexp} into Eq.~\eqref{eq:alpha-WDL} and expanding to first order in $\xi$, we obtain the desired expansion: 
\begin{widetext}
\begin{subequations}\label{}
\begin{numcases}{\alpha(b)=}
\frac{4}{b}\left(1+\frac{\xi}{6}\right)+\frac{15\pi}{4b^2}\left(1+\dfrac{\xi}{3}\right)+\dfrac{128}{3b^3}\left(1+\dfrac{35\xi}{64}\right)+\mathcal{O}\left(\dfrac{1}{b^4}\right),&  $k=2$\,,\label{eq:alpha(b)_2}\\
\frac{4}{b}\left(1+\dfrac{\xi}{2\times3^{k-1}}\right)+\frac{15\pi}{4b^2}\left(1+\dfrac{\xi}{3^{k-1}}\right)+\dfrac{128}{3b^3}\left(1+\dfrac{\xi}{2\times 3^{k-2}}\right)+\mathcal{O}\!\left(\dfrac{1}{b^4}\right),& $k>2$\,.\label{eq:alpha(b)_k}
\end{numcases}
\end{subequations}
\end{widetext}
We note that, similarly to Eq.~\eqref{eq:BH_solB}, Eqs.~\eqref{eq:alpha-WDL} and \eqref{eq:alpha(b)_k} decrease as $k$ increases and $\xi$ becomes smaller, such that the Schwarzschild results are recovered in the limit $\xi\to 0$ for any $k>1$. 
Moreover, we observe that Eq.~\eqref{eq:alpha(b)_k} reproduces the $k=2$ result up to the order $1/b^2$, whereas the $1/b^3$ term exhibits a distinct dependence on $k$.

As an illustrative example, Fig.~\ref{fig:deflection} shows the behavior of the deflection angle for different truncation orders $n$ of the WDL expansion, for $k=2$ and $\xi=0.1$. In this case, the radial distances at which the relative difference with respect to the numerical solution falls below the $\sim1\%$ level are $r\simeq(17.0,10.0,7.9)$ for $n=(3,5,8)$, respectively. As expected, the accuracy of the WDL approximation systematically improves as higher-order terms in the expansion are included. The WDL approximation converges to its numerical prediction for $b\to\infty$.

\subsection{Strong-field regime}
\label{sec:SDL}

We now analyze the SDL occurring when $\alpha(r_0)\to\infty$ or $r_0\to r_m$. In this case, under the change of variable $z=1-r_0/r$ \cite{Li:2026zbo}, Eq.~\eqref{eq:TAA} becomes
\begin{equation}\label{eq:TAA-SDL}
I\left(r_0\right) =\int_0^1 \dd z\, R\left(z, r_0\right) f\left(z, r_0\right) ,
\end{equation}
where ($r=r_0/(1-z)$)
\begin{subequations}
\begin{align}
R\left(z, r_0\right) & =\frac{2 r_0 \sqrt{C_0 A\left(r\right) B\left(r\right)}}{(1-z)^2 C\left(r\right)}\,, \\
f\left(z, r_0\right) & =\left[{A_0-\frac{A\left(r\right) C_0}{C\left(r\right)}}\right]^{-1/2}\,.
\end{align}
\end{subequations}
The function $R(z,r_0)$ is regular for  $r\geq r_0\geq r_m$, while $f(z,r_0)$ is divergent for $z\rightarrow 0$ or $r\rightarrow r_0$. 

The divergent term $f(z,r_0)$ can be treated by performing a Taylor expansion around $z=0$ up to second order:
\begin{equation}
f\left(z, r_0\right) \approx f_0\left(z, r_0\right)\equiv\frac{1}{\sqrt{c_1\left(r_0\right) z+c_2\left(r_0\right) z^2}}\,,
\end{equation}
where
\begin{subequations}
\begin{align}
c_1\left(r_0\right)&=-r_0 A_0^{\prime}+\frac{r_0 A_0 C_0^{\prime}}{C_0}\,,\\ 
c_2\left(r_0\right)&=r_0 \frac{-2 r_0 A_0 C_0^{\prime 2}-C_0^2\left(2 A_0^{\prime}+r_0 A_0^{\prime \prime}\right)}{2 C_0^2}\notag\\
&\quad +\frac{C_0\left[2 r_0 A_0^{\prime} C_0^{\prime}+A_0\left(2 C_0^{\prime}+r_0 C_0^{\prime \prime}\right)\right]}{2 C_0^2}\,.
\end{align}
\end{subequations}

The integral~\eqref{eq:TAA-SDL} can be decomposed into a divergent part $I_D(r_0)$ and a regular part $I_R(r_0)$ as follows \cite{Bozza:2002zj}:
\begin{subequations}
\begin{align}
I\left(r_0\right) & =I_D\left(r_0\right)+I_R\left(r_0\right) ,\\
I_D\left(r_0\right) & =\int_0^1 \dd z\, R\left(0, r_{m}\right) f_0\left(z, r_0\right), \\
I_R\left(r_0\right) & =\int_0^1\dd z\left[R\left(z, r_0\right) f\left(z, r_0\right)-R\left(0, r_{m}\right) f_0\left(z, r_0\right)\right].
\end{align}
\end{subequations}
The regular term $I_R(r_0)$ is generally evaluated numerically, since it does not encode the behavior of $\alpha(r_0)$ in the vicinity of the photon sphere $r_m$. In contrast, the divergent term $I_D(r_0)$ contains the essential information on the SDL and should thus be computed analytically.

In particular, one finds
\begin{align}
I_D(r_0)&= R\left(0, r_{m}\right)\sinh ^{-1}\left[\dfrac{c_1(r_0)}{c_2(r_0)}\right].
\end{align}
We note that, in the limit $r_0\to r_m$ and by means of Eq.~\eqref{eq-ps}, one has $c_1(r_m)\to \infty$, while $c_2(r_m)$ remains finite. Defining $x\equiv c_2(r_0)/c_1(r_0)$, we can therefore perform a Taylor expansion of $\sinh ^{-1}(1/x)$ for $x\to 0$ up to the desired order, namely
\begin{equation}
\sinh ^{-1}\left(\dfrac{1}{x}\right)=\ln \left(\frac{2}{x}\right)+\frac{x^2}{4}+\mathcal{O}\left(x^4\right).    
\end{equation}
Then, we replace $c_2(r_0)$ with its regular value $c_2(r_m)$, whereas $c_1(r_0)$, being divergent, is expanded in a Taylor series around $r_0=r_m$. For all coefficients, we always take into account Eq.~\eqref{eq-ps}. Therefore, we have
\begin{equation}
c_2(r_0)=\sum_{n=1}^\infty c_2^{(n)}(r_m)(r_0-r_m)^n\,,    
\end{equation}
where we report only the leading term
\begin{equation}
c_2^{(1)}(r_m)=\frac{r_m \left(A_m C_m''-C_m A_m''\right)}{C_m}\,.    
\end{equation}

To express the divergent integral in terms of the impact parameter, we first calculate the Taylor expansion of $b-b_m$ in powers of $r_0-r_m$ and then invert the resulting series, obtaining the following expression
\begin{equation}
r_0-r_m=\sum_{n=2}^\infty \mathcal{B}_n(r_m) (b-b_m)^{(n-1)/2}\,,    
\end{equation}
whose leading term is given by
\begin{align}
\mathcal{B}_2(r_m)&= \frac{2 \sqrt[4]{A_m^3 C_m}}{\sqrt{A_m C_m''-C_m A_m''}}\,.   
\end{align}
Implementing these corrections, and after some algebra, one arrives at the following expression at linear order:
\begin{equation}\label{eq:alpha-SDL}
\alpha(b)=-\delta\ln \left(\frac{b}{b_m}-1\right)+\lambda\,,
\end{equation}
where 
\begin{subequations}
\begin{align}
\delta& \equiv \frac{R\left(0, r_m\right)}{2 \sqrt{c_2\left(r_m\right)}}\,,\\ 
\lambda& \equiv\delta\ln\left[r_m^2\left(\frac{C_m''}{C_m}-\frac{A_m''}{A_m}\right)\right]+I_R\left(r_0\right)-\pi\,.
\end{align}
\end{subequations}
We emphasize that all spacetimes of the form~\eqref{eq:general-metric} have this logarithmic behavior of $\alpha(b)$ close to $r_m$ \cite{Bozza:2002zj}. 

\begin{figure}
    \centering
    \includegraphics[width=3.2in]{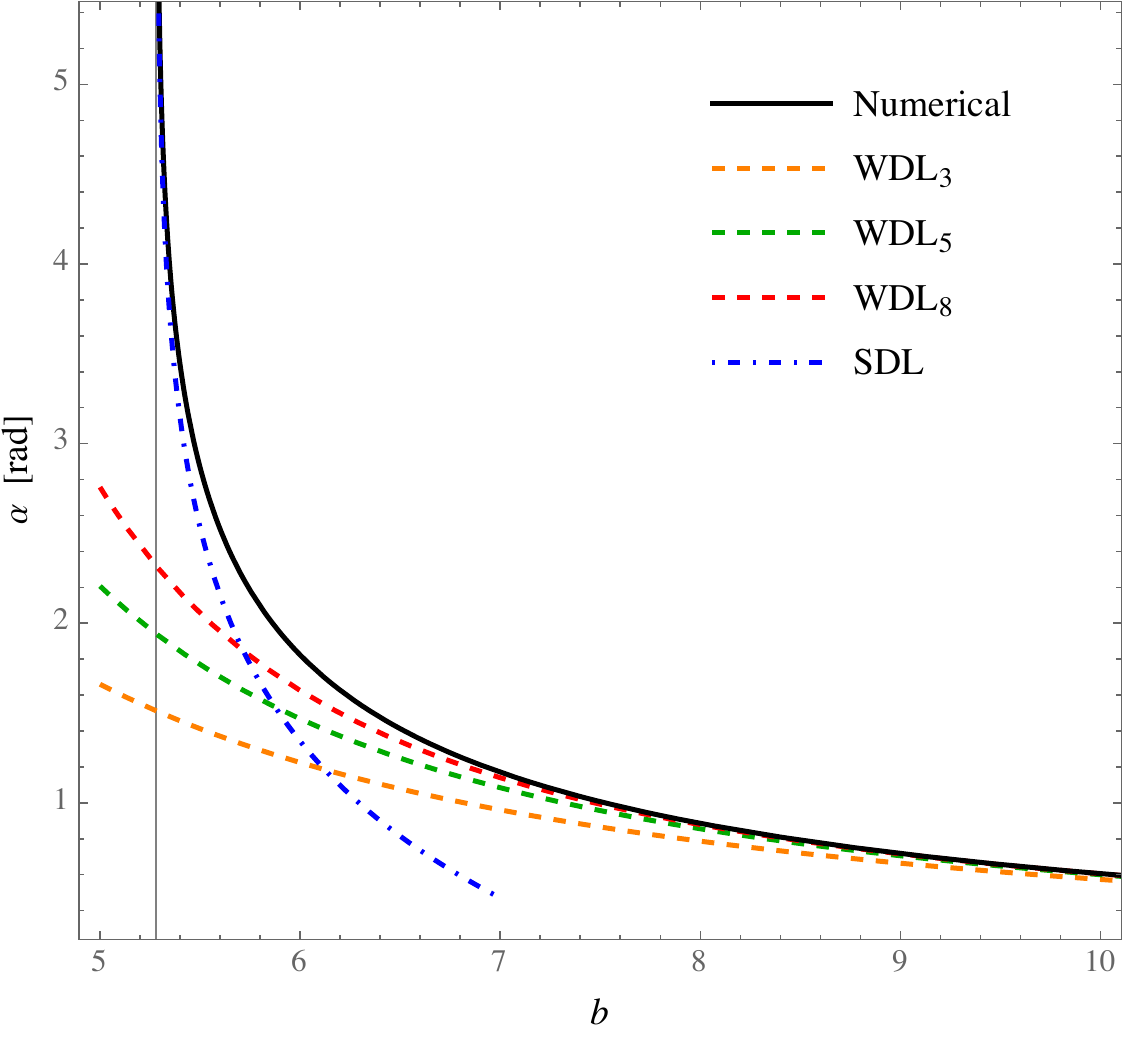}
    \caption{Deflection angle as a function of the impact parameter for $k=2$ and $\xi=0.1$. The solid curve refers to the numerical integration of Eq.~\eqref{eq:DA}. The dashed curves correspond to the expansions~\eqref{eq:alpha(b)_2}-\eqref{eq:alpha(b)_k} for different truncation orders, denoted by $n$. The dash-dotted curve corresponds to Eq.~\eqref{eq:alpha-SDL}. The vertical line marks the critical impact parameter, $b_m=5.28$, associated with the photon sphere $r_m=3.07$.}
    \label{fig:deflection}
\end{figure}

In Fig.~\ref{fig:deflection}, we show the deflection angle in the SDL compared to its numerical prediction for $k=2$ and $\xi=0.1$. We note that the SDL curve converges to the numerical solution for $b\to b_m$, where the deflection angle diverges. We consider only the leading logarithmic term in Eq.~\eqref{eq:alpha-SDL}, as it fully captures the divergent behavior and it is generally sufficient for application purposes.
\section{Astrophysical tests}
\label{sec:applications}

In this section, we examine the gravitational lensing produced by the DD BH in both the WDL (see Sec.~\ref{sec:PPN}) and SDL (see Sec.~\ref{sec:shadow}) via different astrophysical tests. Among these approaches, we restrict our analysis to those satisfying the following criteria: (i) the existence of observational constraints from current observational data, and (ii) effects that are intrinsically associated with BHs rather than other astrophysical systems. 

Finally, in Sec.~\ref{sec:combined}, we complement our analysis by comparing the results achieved in the electromagnetic domain with those previously obtained for the same spacetime \eqref{eq:BH_sol} coming from QNMs in the GW sector \cite{DAgostino:2025yej}.

\subsection{Parametrized post-Newtonian framework}
\label{sec:PPN}

Several astrophysical methods can be exploited to investigate the spacetime~\eqref{eq:BH_sol} in the WDL, including, for example, parametrized post-Newtonian (PPN) formalism, Einstein radius, image position ratios, magnification and flux ratios, and time-delay effects \cite{Keeton:2005jd,Schmidt:2008hc,Will:2014kxa}. 

Given the aforementioned criteria, we focus on the PPN formalism applied to BH systems far from the gravitational source.
The other methods listed above either yield signals too faint to be detected, as their order of magnitude falls below the error systematics of the actual instruments~\cite{Keeton:2005jd}, or probe astrophysical scales different from those relevant to our analysis. 

Specifically, for the spacetime~\eqref{eq:general-metric} with $C(r)=r^2$, the temporal and radial metric components admit the following large-$r$ expansions \cite{Misner:1973prb}:
\begin{subequations}\label{eq:PPN-metric}
\begin{align}
g_{tt}&=-\left(1-\dfrac{2}{r}+\dfrac{2\beta}{r^2}+\cdots\right),\\
g_{rr}&=1+\dfrac{2\gamma}{r}+\cdots, 
\end{align}    
\end{subequations}
where $\gamma$ and $\beta$ are dimensionless PPN parameters encoding deviations from GR, for which $\gamma=\beta=1$. 

The deflection angle can be written in the PPN formalism as\footnote{In standard units, one has $\alpha=(1+\gamma)\frac{2GM}{c^2b}+\mathcal{O}\left(\frac{G^2M^2}{c^4 b^2}\right)$.} \cite{Will:2014kxa,Keeton:2005jd}
\begin{equation} \label{eq:alpha-PPN}
\alpha(b)=\dfrac{2(1+\gamma)}{b}+\mathcal{O}\left(\dfrac{1}{b^2}\right).  
\end{equation}
Comparing this formula with the leading-order expansion of Eq.~\eqref{eq:alpha(b)_k}, we can express $\gamma$ as a function of the DD BH parameters as follows:
\begin{equation}\label{eq:gamma}
\gamma=1+\dfrac{\xi}{3^{k-1}}\,.     
\end{equation}

Although Solar System experiments, such as the Cassini mission, currently provide the tightest bound on the parameter $\gamma$, namely $|\gamma-1|\lesssim10^{-5}$~\cite{Bertotti:2003rm}, such tests probe scales far from the gravitational environment surrounding BHs. Additional estimates of $\gamma$ have also been obtained by applying the PPN formalism to the Einstein radius. In this case, using representative observational uncertainties for stellar-mass BHs from astrometric microlensing measurements~\cite{OGLE:2022gdj}, and for supermassive BHs from stellar-dynamical determinations~\cite{Ghez:2008ms}, one obtains the indicative upper bound $|{\gamma-1}|\lesssim 0.2\,$. 

For our purpose, we instead adopt the more precise astrometric constraints obtained by the GRAVITY collaboration~\cite{Gravity:2019nxk,GRAVITY:2020gka}. Their high-precision monitoring of the S2 star's orbit and its Schwarzschild precession confirms the GR predictions at the Galactic Center with a fractional accuracy of approximately 1\%~\cite{Gainutdinov:2020bbv,Losada:2024wdd}. We can thus reasonably consider the following estimate 
\begin{equation} \label{eq:gamma_bound}
    \gamma= 1.00 \pm0.01\,,
\end{equation}
at the $1\sigma$ confidence level.

\subsection{Black hole shadow}
\label{sec:shadow}

In the SDL as well, several observables can in principle inquire the spacetime geometry, including the angular radius of the photon sphere (or BH shadow), image separation, flux ratios from the photon sphere, and time delays of photons emitted from the photon sphere~\cite{Bozza:2002zj,Tsukamoto:2017fxq}. However, realistic assessments indicate that image separations are generally too small and faint to be resolved by current or near-future instruments \cite{Tsukamoto:2016qro}, flux ratios are second-order in $\xi$ and thus too small to be measured~\cite{Johnson:2019ljv}, and time-delay signatures between multiple relativistic images are similarly challenging to detect with the present observational capabilities~\cite{Gralla:2019xty}.

Therefore, we focus on the BH shadow as a viable probe of the strong-gravity regime. The shadow corresponds to the set of directions on the observer's sky from which photons cannot escape to infinity and are instead captured by the central object; its boundary is determined by unstable photon orbits and is closely related to the photon ring observed in horizon-scale imaging~\cite{EventHorizonTelescope:2019dse,Bronzwaer:2021lzo}. 

In the generic spacetime~\eqref{eq:general-metric}, the BH shadow is a circle that does not depend on the observer's inclination angle with respect to the equatorial plane. The angular radius of the BH shadow is defined as \cite{EventHorizonTelescope:2019dse,Bronzwaer:2021lzo}
\begin{equation} 
\theta_{\rm sh} = \frac{b_m}{D_{\rm OL}}\,.
\label{eq:shadow-radius}
\end{equation}
The photon ring is produced by photons executing one or more nearly circular orbits close to the photon sphere before escaping toward the observer. To leading order, its angular radius coincides with the shadow radius, $
\theta_{\rm ring} \simeq \theta_{\rm sh}$, while its thickness and brightness profile encode higher-order lensing effects and depend sensitively on the emission model and the underlying spacetime geometry~\cite{Bronzwaer:2021lzo}. The angular diameter of the shadow region is $\Theta_{\rm sh} = 2\,\theta_{\rm sh}$, which is generally measured. 

The shadow and photon ring have been imaged by the Event Horizon Telescope (EHT), a global very-long-baseline interferometry array operating at millimeter wavelengths~\cite{EventHorizonTelescope:2019ggy}. At its primary observing frequency of 230 GHz, the EHT achieves an angular resolution of $\theta_{\rm res} \sim 20~\mu{\rm as}$, which is sufficient to resolve horizon-scale structures for nearby supermassive BHs. 

In particular, we consider the angular diameter of the BH shadow, $\Theta_{\rm sh}$, measured by the EHT for both M87${^*}$ and Sgr~${\rm A^*}$. Although the BH solution considered in this work is static and spherically symmetric, we adopt observational constraints derived from systems that are expected to be rotating and, in principle, more accurately described by the Kerr metric. Nevertheless, our methodology is justified by the fact that the angular diameter of the BH shadow depends primarily on the mass-to-distance ratio and only weakly on the spin parameter \cite{Falcke:1999pj}. As a consequence, the average shadow diameter changes only at the level of a few percent even for rapidly spinning BHs, so that EHT measurements provide reliable constraints on our solution.

Therefore, the observed angular scale can be converted into the dimensionless shadow radius in units of BH mass $M$ through the relation
$b_m=(\Theta_{\rm sh}/2)\times (D_{\rm OL}/M)$.

For M87${^*}$, adopting $M=(6.5\pm 0.7)\times 10^9 M_{\odot}$ and $D_{\rm OL}=16.8$ Mpc, with $\Theta_{\rm sh}=(42\pm3$) $\mu$as~\cite{EventHorizonTelescope:2019dse} at $1\sigma$ confidence level, we obtain this result
\begin{equation}
    b_{{m} \rm,M87^*}= 5.50 \pm 0.98\,.
    \label{eq:shadow_M87_bound}
\end{equation}

For Sgr~${\rm A^*}$, using $M=4.0^{+1.0}_{-0.6}\times 10^6 M_{\odot}$, $D_{\rm OL}=8.127$ kpc, and $\Theta_{\rm sh}=(51.8\pm 2.3)$ $\mu$as~\cite{EventHorizonTelescope:2022wkp}, we find
\begin{equation}
    b_{{m}\rm ,SgrA^*}= 5.33 \pm 1.30\,,
    \label{eq:shadow_SgrA_bound}
\end{equation}
at the $1\sigma$ confidence level.

\begin{figure}
    \centering
    \includegraphics[width=3.3in]{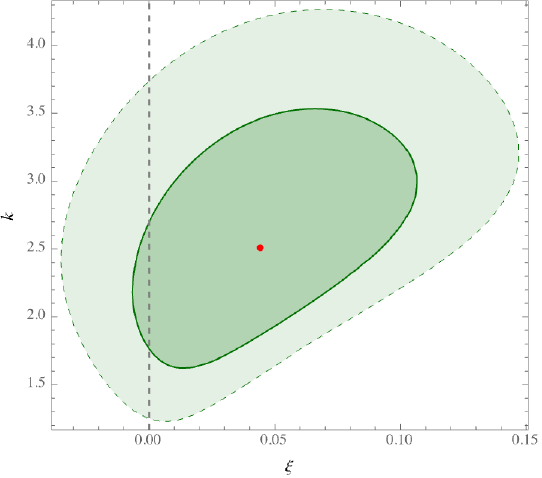}
    \caption{$1\sigma-2\sigma$ confidence regions for the DD BH parameter space, obtained from the joint statistical analysis of the PPN coefficient, the shadow radius, and QNMs measurement constraints. The red point denotes the best-fit values, and the vertical dashed line indicates the GR prediction.}
    \label{fig:combined}
\end{figure}

\subsection{Combined constraints}
\label{sec:combined}

To obtain comprehensive constraints on the DD BH spacetime and quantify potential departures from GR, we complement the estimates discussed above with the bounds obtained from QNMs in our earlier work~\cite{DAgostino:2025yej}. In that analysis, requiring deviations from the Schwarzschild predictions to remain within 10\%, we found the allowed intervals for the DD BH parameters to be $\xi\in(0.015,0.1)$ and $k\in(1,2.84)$. Assuming symmetric $1\sigma$ distributions, these correspond to 
\begin{equation}
    \xi_{\rm QNM}= 0.07\pm 0.04\,, \quad k_{\rm QNM}= 1.7\pm 0.9\,.
    \label{eq:QNM_bounds}
\end{equation}

We therefore perform a joint statistical analysis to reduce the parameter degeneracy and produce tighter constraints on the DD BH parameter space. In particular, we construct a global likelihood function $\mathcal{L}\propto \exp(-\chi^2/2)$, where $\chi^2$ is the sum of statistically independent contributions from the different observables. Specifically, 
\begin{equation}
\chi^2=\chi^2_{E}+\chi^2_{\rm sh}+\chi^2_{\rm QNM}\,,
\end{equation}
where $\chi^2_E$ is relative to the PPN constraint~\eqref{eq:gamma_bound} from the PPN analysis, $\chi_{\rm sh}^2$ to the shadow measurements~\eqref{eq:shadow_M87_bound}-\eqref{eq:shadow_SgrA_bound}, and $\chi_{\rm QNM}^2$ to the QNM bounds~\eqref{eq:QNM_bounds}.
Each $\chi_i$ term is modeled assuming a Gaussian distribution, such that
\begin{equation}
    \chi^2_i(\xi,k)=\sum_j\left[\frac{f_{i,j}^{\rm th}(\xi,k)-f_{i,j}^{\rm obs}}{\sigma_{i,j}}\right]^2\,,
\end{equation}
where $f^{\rm th}_{i,j}$ and $f^{\rm obs}_{i,j}$ are the theoretical predictions and the experimental mean values, respectively, while $\sigma_{i,j}$ are the associated $1\sigma$ uncertainties.

To characterize the parameter space around the best-fit point $\mathbf{p}_{\rm bf}=(\xi_{\rm bf},k_{\rm bf})$, we compute the Hessian matrix 
\begin{equation}
    \mathcal{H}_{ij}=\frac{\partial^2\chi^2}{\partial p_i\partial p_j}\bigg|_{\mathbf{p}_{\rm bf}}\,, \quad i,j\in(\xi,k)\,.
\end{equation}
Close to the minimum, $\chi^2$ can be approximated by its quadratic expansion
\begin{equation}
   \chi^2_{\rm tot}(\mathbf{p}) \approx \chi^2_{\text{min}} + \frac{1}{2} (\mathbf{p} - \mathbf{p}_{\rm bf})^T \mathcal{H}\, (\mathbf{p} - \mathbf{p}_{\rm bf})\,.
\end{equation}
Under the assumptions of Gaussian errors, the Hessian is related to the Fisher information matrix, $\mathcal{F}$, through $\mathcal{F}=\frac{1}{2}\mathcal{H}$ \cite{Cutler:1994ys}.
The covariance matrix is then given by
\begin{equation}
\mathcal{C} = \mathcal{F}^{-1} = 
\begin{pmatrix} 
\sigma_{\xi}^2 & \sigma_{\xi k} \\ 
\sigma_{k\xi} & \sigma_k^2 
\end{pmatrix},
\end{equation}
where the diagonal elements give the marginal variances, whereas the off-diagonal terms quantify the parameter correlation.

The Fisher matrix method is widely used in parameter estimation and forecasting analyses across different astrophysical and cosmological contexts (see, e.g., Refs.~\cite{Tegmark:1997rp,Seo:2003pu,Yunes:2009ke,Weinberg:2013agg,Bonilla:2019mbm}). In the present case, the number of observational inputs is relatively small, and the parameter space is two-dimensional, which makes a full Markov Chain Monte Carlo exploration unnecessary. The Fisher-matrix approximation thus provides an efficient and robust way to quantify parameter degeneracies and estimate confidence regions through the local curvature of the $\chi^2$ surface. While a fully Bayesian approach would provide a more complete characterization of the posterior likelihood, it generally requires significantly greater computational resources involving explicit implementations of the noise~\cite{Vallisneri:2007ev}.
Accordingly, the present analysis provides a first, quantitative assessment of the viable region in the DD BH parameter space, offering guidance for a complete Bayesian inference in view of future larger datasets.

Confidence regions in the $(\xi,k)$ plane are determined by the level sets of the $\chi^2$ surface satisfying the condition
$\Delta\chi^2=\chi^2-\chi^2_{\rm min}\leq\Delta\chi^2_{\alpha}$, where $\alpha$ defines the significance level. For two degrees of freedom, the $1\sigma$ ($\sim68\%$) and $2\sigma$ ($\sim 95\%$) confidence regions correspond to $\Delta\chi^2=2.30$ and $\Delta\chi^2=6.18$, respectively.

The results of our combined analysis are shown in Fig.~\ref{fig:combined}, which finally allows us to obtain the $1\sigma$ estimates:
\begin{equation}
\left\{
\begin{aligned}
    \xi &= 0.044 \pm 0.039\,, \\
    k &= 2.51 \pm 0.64\,.
\end{aligned}
\right.
\end{equation}
The statistical consistency with the Schwarzschild metric can be quantified through the difference
\begin{equation}
 \Delta\chi^2_{\rm GR} = \min\limits_{k}\chi^2(\xi=0, k)- \chi^2_{\rm min}\,.
\end{equation}
In our case, marginalizing over $k$, the resulting statistics follows a $\chi^2$ distribution with one degree of freedom. We therefore obtain
\begin{equation}
    \Delta\chi^2_{\rm GR}=\frac{\xi_{\rm bf}^2}{\sigma_{\xi}^2}=1.28\,,
\end{equation}
corresponding to a $1.13\sigma$ consistency with GR.

\section{Conclusions}
\label{sec:conclusions}

In this work, we investigated the gravitational lensing produced by static and spherically symmetric DD BH solutions within the revised Deser--Woodard nonlocal gravity model. Starting from the modified spacetime geometry, which introduces a perturbation parameter $\xi$ and an exponent $k$ representing nonlocal corrections to the Schwarzschild metric, we derived the light deflection angle in both the WDL and SDL approximations.

In the WDL, we provided analytical expansions of the bending angle as a function of the impact parameter, highlighting its explicit dependence on the nonlocal parameters. In the SDL, associated with photons completing multiple loops near the photon sphere, we verified that the deflection angle exhibits the universal logarithmic divergence characteristic of static and spherically symmetric spacetimes, and we computed the corresponding coefficients for the DD BH case.

To bridge the gap between theory and observations, we applied our analytical results to measurable astrophysical quantities. 
In particular, we considered current estimates of the PPN parameters by GRAVITY from the orbit of the S2 star. We also used the high-resolution bounds on the BH shadow angular diameter obtained by the EHT for $\rm{Sgr~A}^*$ and $\rm{M87}^*$.
Finally, we combined such constraints with those previously derived from the QNMs in the GW sector. We thus performed a joint likelihood analysis based on the Fisher information matrix to find tighter constraints on the DD BH parameter space. In doing so, we obtained $\xi=0.044 \pm 0.039$ and $k=2.51 \pm 0.64$, indicating consistency with GR at the $1.13\sigma$ confidence level. 

Our results show that DD BH solutions provide an observationally viable framework for testing deviations from GR. Future advances in high-precision astrometry and next-generation horizon-scale imaging will be crucial for tightening current bounds and elucidating potential nonlocal effects in the strong-field regime. 
Further progress may be achieved by improving the statistical analysis through more accurate parameter measurements and by incorporating additional independent constraints from complementary multimessenger probes, including electromagnetic and GW observations. Such a systematic approach would enable more stringent tests of the standard gravitational paradigm.

\acknowledgements
The authors would like to thank Haida Li and Xiangdong Zhang for useful discussions.
R.D. acknowledges the financial support of  Istituto Nazionale di Fisica Nucleare (INFN), Sezione di Roma 1, \textit{esperimento} Euclid. 
V.D.F. wishes to thank Gruppo Nazionale di Fisica Matematica of Istituto Nazionale di Alta Matematica (INDAM) for support. V.D.F. acknowledges the support of INFN, Sezione di Napoli, {\it iniziativa specifica} TEONGRAV. V.D.F. dedicates this work to the memory of Prof. Pavel Bakala and Prof. Maurizio Falanga.

\bibliography{references}

\end{document}